\documentclass[preprint,preprintnumbers,amsmath,amssymb]{revtex4}

\usepackage{graphicx}
\usepackage{dcolumn}
\usepackage{bm}

\begin{document}

\preprint{  }

\title{SmB$_6$: Topological insulator  or semiconductor with\\
 valence-fluctuation induced hopping transport?}

\author{I. Batko}%
 \email{batko@saske.sk}  
 \affiliation{Institute  of  Experimental  Physics,
 Slovak   Academy  of Sciences, Watsonova 47,
 040~01~Ko\v {s}ice, Slovakia}

\author{M. Batkova }   
 \email{batkova@saske.sk}
 \affiliation{Institute  of  Experimental  Physics,
 Slovak   Academy  of Sciences, Watsonova 47,
 040~01~Ko\v {s}ice, Slovakia}           

\date{\today}

\begin{abstract}
We advert to the fact that presence of valence fluctuations (VFs)
   in semiconductors with in-gap impurity bands unconditionally leads to 
   dynamical changes (fluctuations) of energies of localized impurity states.   
%
%
We provide arguments that in the impurity subnetwork consisting of centers 
		having energy levels
		fluctuating around the Fermi energy there exist favorable conditions for hops
		from occupied states to empty states of less energy. 
Consequently, we propose original {\em valence-fluctuation induced} hopping mechanism 
		as a new possibility to explain 
		unusual metallic-like conduction
		of SmB$_6$ and other Kondo insulators 
		experimentally observed	at lowest temperatures.
Interestingly, the proposed mechanism infers enhanced metallic-like 
		surface conductivity of SmB$_6$, 
		what resembles a characteristic property of topological insulator, 
		and is in agreement with experimental observations attempting to prove
		the existence of topologically protected surface state in SmB$_6$.    
		
\end{abstract}


\maketitle

SmB$_6$ belongs to the class of materials known as heavy-fermion semiconductors, 
		or alternatively as Kondo insulators \cite{Riseborough00}.
These materials are characterized by electronic properties, 
		which at high temperatures are associated 
		with a set of independent localized ($f$) magnetic moments 
		interacting with a sea of conduction electrons, 
		while at low temperatures their electronic properties 
		resemble those of narrow gap semiconductors
		\cite{Riseborough00,Wachter93,Antonov2002}.
Specifically for SmB$_6$, a small electronic gap ($\Delta = 2 - 30$~K) 
		was detected by many experimental techniques,
		e.g. optical conductivity \cite{Travaglini84}, infrared absorption \cite{Namba1993,Ohta1991}, 
		inelastic neutron 	scattering \cite{Alekseev1993},
		electron tunneling \cite{Frankowski1982,Batkova2008,Batkova2006}, 
		and electrical transport measurements \cite{Batkova2006,Nickerson1971,Allen79, Batko93, Cooley95,
		Flachbart01pb, Gabani01}.
A fundamental phenomenon that governs physical properties of SmB$_6$ is 
		intermediate valence (or homogeneously mixed valence)
		of Sm ions that fluctuate
		between Sm$^{2+}$ ($4f^{6}$) and Sm$^{3+}$ ($4f^{5}5d^{1}$) configurations \cite{Wachter93}.
The Sm$^{2+}$~:~Sm$^{3+}$ ratio at room temperature is about 4~:~6 
		and varies with temperature weakly,  
		as established by measurements of lattice constant \cite{Tarascon1980}, x-ray 									
		absorption \cite{Tarascon1980}, and M\"ossbauer studies \cite{Cohen1970}.
Studies of spin and charge fluctuations
		due to Sm$^{2+}$ $\rightleftharpoons$ Sm$^{3+}+d$ fluctuations
		indicate fluctuation rates in the range of phonon frequencies;
the spin fluctuation time obtained via $^{10}$B and $^{11}$B NMR between 2 and 300~K 
		yields a spin fluctuation rate
		$\tau_{sf}^{-1} \sim 10^{13}$~s$^{-1}$ \cite{Pena1981}, 
		while phonon spectroscopy studies indicate charge fluctuations
		between 200~cm$^{-1}$ and 650~cm$^{-1}$ \cite{Zirngiebl86,Mock86},
		what corresponds to the characteristic charge fluctuation rate, $\tau_{cf}^{-1}$, 
		between $6 \times 10^{12}$~s$^{-1}$ and $2 \times 10^{13}$~s$^{-1}$.

Although SmB$_6$ and other Kondo insulators have been 
		intensively studied for more than  four decades, 
		many fundamental aspects of the microscopic description of the mixed valence ground state
		and the nature of the VFs are still under discussion
		\cite{Antonov2002}.
One of the principal questions is whether Kondo insulating materials actually 
		are true insulators at low temperatures
		or whether small conduction-electron carrier concentration is present
		\cite{Antonov2002,Fisk1996}.
Answering this question is especially complicated in the case of SmB$_6$ 
		due to a fundamental problem  to interpret consistently
		the electrical resistivity/conductivity behavior at lowest temperatures.

Electrical resistivity measurements of SmB$_6$ show a large
		resistivity increase at decreasing temperature below 50~K,
		and saturation  
	  at very high residual value $\rho _0$  
		at lowest temperatures \cite{Allen79, Batko93, Cooley95,
		Flachbart01pb, Gabani01}.
 The corresponding electrical conductivity $\sigma(T)=1/\rho(T)$
		in the "saturation region"
		is metallic-like and reveals a temperature non-activated (metallic) channel in $\sigma(T)$, 
		which becomes dominating below about 3~K 
		\cite{Batko93, Cooley95, Flachbart01pb, Gabani01, Gabani03}. 
This residual conductivity can be attributed 
		to in-gap states, which were confirmed by other probes 
		\cite{Namba1993,Ohta1991,Caldwell07,Miyazaki2012}.  
However, 
		detailed simultaneous studies of the resistivity and Hall resistivity 
		performed on highest quality SmB$_6$ samples indicate  
   	that very high value of $\rho_0$ would require a superunitary 
		scattering \cite{Allen79,Cooley95} with 
		unphysically high concentration of scattering centers 
		(at least 80 per unit cell \cite{Cooley95}).  
Taking into account the Mott-Ioffe-Regel viewpoint, 
		which claims that the conventional Boltzmann transport theory becomes meaningless
		when the characteristic mean free path 
		of the itinerant conduction	electrons becomes comparable to,
		or less than the interatomic 
		spacing \cite{Ioffe60,MottDavis81,Edwards98},
the very high value of the residual resistivity in SmB$_6$ 
		cannot be attributed to metallic conductivity mediated by 
		itinerant electrons.
Thus, some kind of hopping transport
	  is required to describe the conductivity of SmB$_6$ at lowest temperatures.
However, hopping-type conduction within the framework of present theoretical models 
		requires a temperature activated conductivity and an insulating 
		ground state, what is in qualitative disagreement 
			with the above mentioned experimental indications 
			\cite{Batko93, Cooley95,Flachbart01pb, Gabani01, Gabani03}. 
Hence, the origin of the residual resistivity/conductivity
			in SmB$_6$ can not be consistently explained considering any 
			known mechanism of electrical conductivity in metals 
			or  semiconductors if a homogeneous bulk conduction is considered
			to be responsible for the observed electrical conductivity.
Possible solution of this mystery represents recent proposition that SmB$_6$ 
		is a topological Kondo insulator \cite{Dzero2010,Dzero2012,Lu2013},
		so that topologically protected metallic surface states can be responsible 
		for the anomalously high residual resistivity. 
Although several studies indeed indicate metallic surface transport in SmB$_6$ 
		\cite{Kim2012,Wolgast2013,Kim2013},
		the question about nature of the surface states in SmB$_6$ remains open. 
Moreover, new points of view on the origin of the surface enhanced conductivity in SmB$_6$
		were recently presented,
		e.g. polarity driven surface metalicity in SmB$_6$ \cite{Zhu2013} 
	or coexistence of non-trivial 2D surface state and trivial surface layer  
		 \cite{Chen2013}.

The purpose of this paper is to show that there exists also other possibility to explain 
		mysterious residual resistivity of SmB$_6$ within the concept of hopping transport,
		if effect of VFs of samarium ions on the energy of in-gap impurity
		states is taken into consideration. 
Our explanation is based on the fact that  Sm$^{2+}$ $\rightleftharpoons$ Sm$^{3+}+d$ fluctuations of valence
		are {\em intrinsically} associated with 
		fluctuations of charge, magnetic moments and ionic radii of samarium ions.
Because all these parameters influence the interaction energy
		of an impurity center with the surrounding lattice
		(thus influencing the energy of corresponding localized impurity state) 
		it has to be concluded that
 		fluctuations of the mentioned parameters will {\em unconditionally} cause 
 		fluctuations of  energies of localized impurity states 
 		in the impurity band of SmB$_6$. 
As we show in this paper,	
		just fluctuations 
		of energies of impurity states 
		''driven" by VFs  may
		give rise to a new type of hopping mechanism, which is energetically
		more favorable at lowest temperatures 
		than the well-known variable-range hopping (VRH).
Hopping probability in this new hopping mechanism is governed
		by dynamics of valence fluctuations and
		by a characteristic distance between hopping centers,
what
		in case of increased concentration of hopping centers
		in the surface layer yields
		correspondingly increased hop probability in the surface layer. 
The proposed hopping mechanism can be 
		responsible for metallic-like  conduction of SmB$_6$ 
		at lowest temperatures, 
		as follows from the analysis and the discussion presented below.
%

	\begin{figure}
			\center{
				\resizebox{0.95\columnwidth}{!}{%
  				\includegraphics{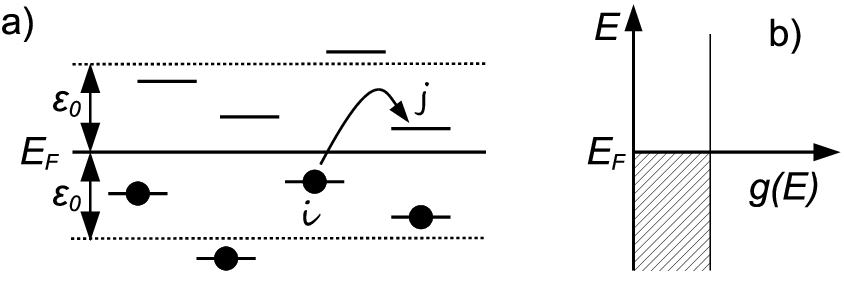}
        }
    	}
 			\caption{Schematic depiction of states from the impurity band 
 			with energies lying in the vicinity of the 
 			Fermi energy in case of a classic semiconductor (a), 
 			and corresponding DOS diagram (b). 
  		}
		\label{fig1}
		\end{figure}

Let us start with a model case of a classic  semiconductor
  having a simple cubic crystal lattice with the lattice parameter $a$,
  containing metal ions, $Me$, 
  positioned in the centers of the elementary cells.
Let's also suppose that metal ions are in two different valency states, say $Me^{2+}$ and $Me^{3+}$, 	
	where $Me^{2+}$ and $Me^{3+}$ ions are randomly distributed over the metal ion positions, 
	while their valences are ``static'', i.e. they do not change in time.
%
%
In addition, let us suppose that 
	(i) the considered semiconductor contains 
	localized donor impurity states that form IB in its forbidden gap,	
	(ii) the Fermi energy, $E_F$, lies in the IB, 
	and 
	(iii) the IB in the vicinity of the Fermi energy  
	is characterized 	by a constant density-of-states (DOS) function, $g(E)$. 
The supposed situation is similar to that well-known
	for doped (classic) semiconductors 
	as schematically depicted in Fig. \ref{fig1}
	 (see e.g. \cite{MottDavis81, Shklovskij84}).
So, it is reasonable to suppose that  
		electrical conduction in such system 
		is temperature activated at low temperatures 
		and can be adequately described
		by hopping process utilizing the concept of Miller-Abrahams network \cite{Miller1960},
		where randomly distributed impurities form a resistance network, 
		having the nearest neighboring impurities at a typical distance 	$R>>a$
		\cite{MottDavis81, Shklovskij84, Mott1968}.
However, as noted for the first time by Mott \cite{Mott1968},
		the activation energy needed to hop
		should in principle drop to zero at lowest temperatures, 
		from the following reasons. 
If the electron hops a much bigger distance (say $pR$ with $p>>1$)
		it has a choice of $p^3$ as many sites and may expect to find some 
		with an energy nearer to its own, say with energy difference $W_{D}/p^3$ \cite{Mott1968},
where $W_D$ is activation energy defined by Miller and Abrahams \cite{Miller1960} 
		as the average energy
		needed to hop to nearest neighboring impurities at distance $R$.
Hopping probability for hops to distance $pR$  
		may be written as proportional to \cite{Mott1968}  
\begin{equation} 		
 	e^{(-2\alpha pR - W_D/p^3kT)},
 		\label{Mott-phop} 
\end{equation}				
where $\alpha^{-1}$ is a localization length.
Expression (\ref{Mott-phop}) has a maximum when \cite{Mott1968}
\begin{equation} 		
 	p^4 = 3 W_D / 2\alpha R k T.
 		\label{Mott-pmax} 
\end{equation}				
As concluded by Mott,  if (\ref{Mott-pmax}) gives a value of $p$ greater than unity, 
		the hopping probability is no longer proportional to $ e^{-W_D/kT}$
		but obeys the proportionality 
\begin{equation} 		
 	P_{Mott} \propto	e^{-(T_{0}/T)^{1/4}},
 		\label{Mott-vrhc} 
\end{equation}
where  $T_{0}= const.(\alpha R)^{3}W_{D}/k$ \cite{Mott1968}, 
		and $P_{Mott}$ is corresponding hopping probability. 
 Thus, according to the original Mott's analysis 
 		and other more detailed works
 			\cite{Mott1968,MottDavis81,Ambegaokar71,Shklovskij84},
			conductivity due to electron hops of our model system
			containing $Me^{2+}$ and $Me^{3+}$ ions  
			can be written as proportional to $e^{-(T_0/T)^{1/4}}$
			at lowest temperatures,
while a typical activation energy of a hop,
			$\epsilon_0 \approx W_{D}/p^{3}$,
			decreases with temperature 
			as $T^{3/4}$.
It can be said that conduction    
 		arises from the phonon assisted hopping 
 		with a typical activation energy $\epsilon_0$
 		in a so-called optimal band,
 		where the optimal band is defined as a narrow energy interval of the width 	
 		$2\epsilon_{0}$  centered at the Fermi level (see Fig. \ref{fig1}a)
		with concentration of impurities \cite{Shklovskij84}
\begin{equation} 		
 		N(\epsilon_{0})=2\epsilon_{0}g(E_F).
 		\label{OB-conc} 
\end{equation}			
Because concentration of impurities in the optimal band decreases 
	with temperature and approaches zero for $T \to 0$
	[$N(\epsilon_{0})\propto \epsilon_{0}\propto T^{3/4}$],
   a typical hop distance ($\propto T^{-1/4}$) goes to infinity for $T \to 0$,
	 resulting in insulating ground state. 

\begin{figure}
				\center{
					\resizebox{0.95\columnwidth}{!}{%
  			\includegraphics{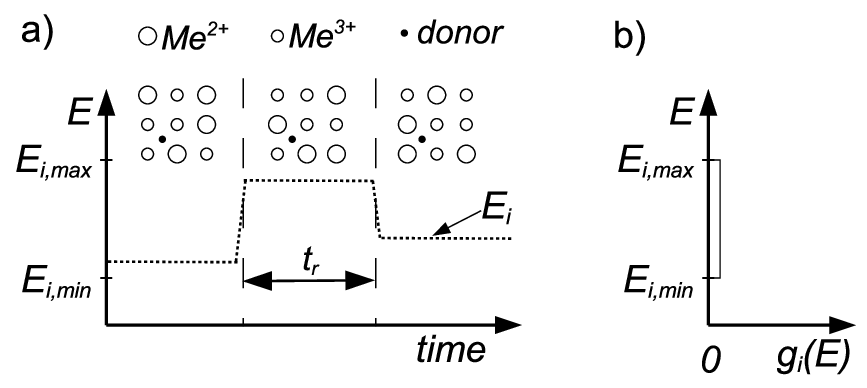}
            }
    }
 			\caption{
Time evolution of donor energy $E_i$  
  		due to rearrangement process of metallic ions (a) 
  		and corresponding time-averaged partial DOS (b)
  		of the donor state. 
  		Rearrangement process in this example is realized as
  		clockwise rotation (90 degrees) of metallic ions
  		around ''central'' $Me^{3+}$ ion. 
 }
 \label{fig2}
\end{figure}

Now, let us ``turn on" a {\em hypothetical} rearrangement process (RP) 
			in the considered semiconductor that causes repetitious
			changes in distribution of $Me^{2+}$ and $Me^{3+}$ ions on metal ion positions
			with a characteristic time constant, $t_r$.
We consider only such rearrangements, where 
			the number of $Me^{2+}$ and $Me^{3+}$ 	
			ions is not changed (i.e. RP is realized via position interchanges of  some $Me^{2+}$ and $Me^{3+}$ ions) 
			and the macroscopic physical properties of the system remain unchanged. 
(The situation is depicted in Fig. 2.)
Due to different physical properties of $Me^{2+}$ and $Me^{3+}$ ions,
		 changes in space distribution of 
			 $Me^{2+}$/$Me^{3+}$ ions surrounding an impurity 
			 will correspondingly change interaction of the impurity with the surrounding lattice.  
For instance, ``replacement" of $Me^{2+}$ by $Me^{3+}$ (with relatively smaller radius)
			will  decrease  local chemical pressure acting on the neighboring impurity center 
			and increase the attractive Coulomb interaction between the (donor) impurity center  
			and the metal ion. 
Therefore, RP-caused 
		different redistributions of $Me^{2+}$/$Me^{3+}$ ions 
		surrounding an impurity center $i$ will 
		result in different  energy
		of the corresponding localized impurity state, $E_i$,	
		thus  $E_i$ will change with a characteristic time constant $t_r$
		of the RP, as schematically shown in Fig.~\ref{fig2}a.
Performing all possible rearrangements as defined above 
		(i.e. all possible distributions of $Me^{2+}$/$Me^{3+}$ ions 
		that do not change macroscopic/bulk physical properties of the material), 
		the energy $E_i$ will change within the interval expressed by the inequality
				\begin{equation}
					E_{i, min} \leq E_i \leq E_{i, max}.
						\label{Eminmax} 
				\end{equation}
Here, $E_{i, min}$ and $E_{i, max}$ ($E_{i, min} < E_{i, max}$) are the minimum and maximum possible 
			energy of the donor, respectively. 
So that, as follows from the arguments above, the energy of localized impurity/donor 
    state changes with every rearrangement, and lies in energy interval 
		of non-zero width 
					\begin{equation}
						\Delta E_{i} = E_{i, max} -  E_{i, min} > 0.
							\label{bandwidth}
					\end{equation}
In other words, the localized impurity/donor state is not longer characterized by energy constant in time 
    (like in semiconductor with absence of RP as depicted in Fig. \ref{fig1}a),
    but by energy interval  (resembling a narrow band), e.g. like shown in Fig.~\ref{fig2}b.
%
Probability of finding the localized impurity/donor 
		state $i$ at the energy $E$ is  characterized by 
		partial-DOS function $g_i(E)$,
		which is zero outside the interval defined by inequality (\ref{Eminmax}),
		but might have rather ``complex shape" within this interval.
For purposes of this paper $g_i(E)$ is  considered to be constant within 
		the interval defined by inequality (\ref{Eminmax}), as depicted in Fig.~\ref{fig2}b. 

RP-caused changes of energy of localized impurity states in time have principal impact on hopping process.
According to our original supposition,
		 $E_F$ lies in the IB which is characterized by a constant DOS,
		 and we consider only such redistributions of $Me^{2+}$/$Me^{3+}$ ions
		 that do not change physical characteristics of the system 
		 (i.e. DOS of the whole macroscopic system remains 
		 unchanged). 
Thus, there have to exist impurity centers $i^*$ 
		with corresponding energy if localized states $E_{i^{*}}$ satisfying the condition
		$E_{i^{*}, min} < E_{i^{*}} < E_{i^{*}, max}$,
		such that
\begin{equation}
		E_{i^{*}, min} < E_F < E_{i^{*}, max}.
		\label{Ef-state}
\end{equation}
Assuming for simplicity  that impurity states are characterized by 
		an equal/typical width of the energy interval
$\Delta E_{i} \approx E_{0}$, 
then energies of impurity centers $E_{i^{*}}$ which due to RP 
		can be less, as well as greater than $E_F$, satisfy the inequality
\begin{equation}
				E_F - E_{0} < E_{i^{*}} < E_F + E_{0}.
				\label{Ef-band}
\end{equation}
The concentration of such centers is 
	\begin{equation}
	N^{*}(E_{0}) = E_{0}g(E_F).
	\label{Ef-conc}
	\end{equation} 
Correspondingly, a typical distance between two nearest centers is
\begin{equation}
R^{*}=[N^{*}(E_{0})]^{-1/3} = [E_{0}g(E_F)]^{-1/3}.
	\label{Ef-distance}
	\end{equation}

Impurity centers defined by (\ref{Ef-band}) have a unique property;
there exists a non-zero probability 
		that some occupied  donor levels will shift  due to RP
		from the region below $E_F$ to the region above $E_F$,
		and analogously,
	 some empty  donor levels from the region above $E_F$ will shift under $E_F$.
Such energy changes	represent in fact non-equilibrium excitations
		(driven by RP)
		that have to be consequently brought into the equilibrium state,
what happens via 
		electron hops of ``zero activation energy'' from occupied  states to empty states of less energy. 
		This in fact, indicates a possibility of 
		new hopping mechanism, which should be realized via temperature non-activated hops (tunneling) 
		that have to dominate at lowest temperatures.
	 	Let us analyze the situation in more details.

The intrinsic transition rate $\gamma_{ij}$ for an electron hopping from a site 
		$i$ with energy $E_i$ to an empty site $j$ with energy $E_j$
	  in the simplest case, when $kT$ is small compared to
		$|E_j-E_i|$, and $|E_j-E_i|$ is of the order of the Debye energy or smaller, 
		is well approximated by the ``quantum-limit'' hopping formula \cite{Ambegaokar71} 
\begin{eqnarray}
		\gamma_{ij} =	\gamma_{0}e^{-2\alpha R_{ij} - (E_j-E_i)/kT} \mbox{ for } E_j > E_i\\
		\gamma_{ij} = \gamma_{0}e^{-2\alpha R_{ij}} \mbox{ for } E_j < E_i ,
				\label{TNAH}
	\end{eqnarray} 
		where $\gamma_{0}$ is a constant as defined elsewhere \cite{Ambegaokar71} 
		and $R_{ij}$ is the distance between the centers $i$ and $j$.
Bidirectional hopping process 
		between the centers $i$ and $j$ ($i \to j \to i \to j \dots$),  		 
in a classic semiconductor (without RP) 
		unconditionally requires thermal activation,
		because bidirectional hopping always includes also hops to empty states of higher energy 
		[see (11) and Fig. \ref{fig1}a].
For instance, if the first hop is one 
		to a site of less energy,
		it does not require thermal activation
		[as follows from (\ref{TNAH})].
However, the following hop in opposite direction 
		has to be one to a state of higher energy (see Fig. 1a),
	  so it requires temperature activation 
		[as follows from (11)].
		This implicates that bidirectional hopping process 
		is conditioned by energy (phonon) absorption. 
%
%
[In fact, substitution of $R_{ij}$ and $(E_j - E_i)$ in (11) by 
		typical hop distance $pR$ and activation energy $W_{D}/p^{3}$, respectively,  
		yields the same exponential term as (\ref{Mott-phop}), 
		which in accordance with Mott's original deviation \cite{Mott1968}
		(as summarized in the beginning) leads to VRH-conductivity 
		proportional to  $e^{-(T_0/T)^{1/4}}$.]


		%
\begin{figure}
				\center{
					\resizebox{0.95\columnwidth}{!}{%
  			\includegraphics{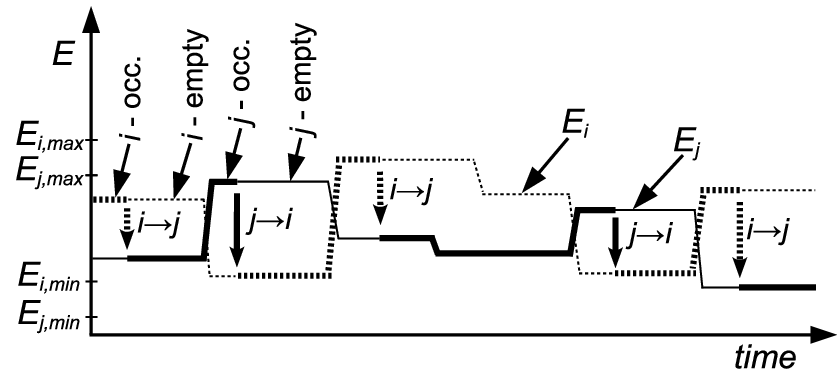}
            }
    }
 			\caption{
Schematic depiction of bidirectional hopping between states $i$ and $j$
	due to rearrangement processes.
Time evolution of energy of $i$ and $j$ state is represented by 
	dashed and solid line respectively, while occupation of the state is 
	represented by the line thickness (thick - occupied, thin - empty).
Hops from $i$ and $j$ states are indicated by vertical arrows with dashed and
	solid lines, respectively.   
 }
 	\label{fig3}
\end{figure}

The situation is qualitatively  different in a semiconductor, 
		where RP of metal ions $Me^{2+}/Me^{3+}$
		takes place, thus energies of impurity states change in time.  
Let's consider two impurity states ($i$, $j$) that have non-zero overlap 
		of their energy intervals 
		(say $E_{j,min} < E_{i,min} < E_{j,max} < E_{i,max}$ like shown in Fig. \ref{fig3}),
		and RP causing repetitive changes of mutual positions of energy 
		levels $E_i$ and $E_j$.  	
If 	the first hop (say $i \to j$) is from an impurity site of (temporarily) higher energy
		to one of less energy (i.e. $E_i > E_j$) then it does not require thermal activation
		(see Fig. \ref{fig3}).
However, there exist {\em two} possibilities 
		for the hop in the opposite direction ($j \to i$):
		(i) the hop will require thermal activation if performed 
		during the time period when $E_j < E_i$, 
		or (ii) the hop will be performed without thermal activation 
	 {\em after} at least one rearrangement of metal ions $Me^{2+}/Me^{3+}$ 
		causing mutual change of relative positions of energy levels $E_i$ and  $E_j$,
		when  $E_j$ becomes (temporarily) greater than $E_i$, as schematically depicted in Fig. {\ref{fig3}. 
Introducing a	parameter $\nu_{ij}$ ($0 < \nu_{ij} <1$) 
		denoting the probability 
		that an occupied state $i$ occurs due to RP at
		(temporarily) higher energy level than an empty state $j$,
		then the intrinsic transition rate $\gamma_{ij}^{non-act}$ 
		for thermally non-activated electron hop from site $i$ to site $j$
		with (temporarily) less energy will be 
\begin{equation}
		\gamma_{ij}^{non-act} = \nu_{ij}\gamma_{0}e^{-2\alpha R_{ij}},
		\label{RP-na}
\end{equation}
		while the intrinsic transition rate $\gamma_{ij}^{act}$ 
		for thermally activated electron hop from site $i$ to site $j$
		(with temporarily higher energy) will be
\begin{equation}
		\gamma_{ij}^{act} =	(1-\nu_{ij})\gamma_{0}e^{-2\alpha R_{ij} - (E_j-E_i)/kT}.
\end{equation}
If the ratio	
\begin{equation}
		\beta_{ij} = \frac{\gamma_{ij}^{non-act}}{\gamma_{ij}^{act}}
= \frac{\nu_{ij}}{(1-\nu_{ij})e^{-(E_j-E_i)/kT}}
	\label{Bij}
\end{equation}
			is greater than unity, the predominating electron hops between impurity states $i$ and $j$
			are not longer thermally activated, 
			but thermally non-activated hops to states with less energy that are ``induced'' by 
			RP.
Because for every non-zero $\nu_{ij}$ there exists a temperature $T_{ij}$ providing  $\beta_{ij} >> 1$ for $T<T_{ij}$,  
intrinsic transition rate $\gamma_{ij}$ of the considered 
		bidirectional hopping process   at lowest temperatures
	  can be in {\em both directions} approximated by
\begin{equation}
 \gamma_{ij} \approx \nu_{ij}\gamma_{0}e^{-2\alpha R_{ij}}.
 \label{Gij-tnat}
\end{equation}
%
%
However,
this practically means that for $T \to 0$ there is no need for electron to hop to empty state of higher energy,
		because it is energetically more favorable 
		to "wait'' for a ''proper" rearrangement of $Me^{2+}/Me^{3+}$ ions causing
		that empty state of less energy occurs
		(just as it is schematically shown in Fig. \ref{fig3}).
%
%
%
			
Considering subnetwork of impurities satisfying the inequality (\ref{Ef-band})
			we define $\nu^{*}_{rp}$ as the time-averaged
			probability per unit volume that for a {\em temporarily} occupied state belonging to the 
			subnetwork, an empty state from the subnetwork will appear at lower energy as a consequence of RP. 
Considering electron hops to sites within the distance $pR^{*}$, 
		where $p\ge1$ and $R^{*}$ is given by  Eq. (\ref{Ef-distance}),
		electron has choice of $p^3$ as many sites to find an empty state of less energy.
		Probability of such RP-induced hop, $P^{*}_{rp}$, can be expressed as 
\begin{equation}
	P^{*}_{rp} \propto p^3 \nu^{*}_{rp} e^{-2\alpha pR^{*}},
	\label{phopRP}
\end{equation}
and has maximum for  $p=3 / 2 \alpha R^{*}$. 
If value  $3 / 2 \alpha R^{*} \ge 1$, 
		maximum probability of RP-induced hop 
		can be (omitting numerical factors) expressed as
\begin{equation}
		P^{*}_{rp} \propto  \nu^{*}_{rp}/(\alpha R^{*})^3.
\end{equation}
However, estimation of $\alpha R$ in classic doped semiconductors ($R^{*} \ge R>>a$) \cite{Pol-Sklov1991}
		indicates  values $3 / 2 \alpha R^{*} < 1$. 
Therefore, predominated hops should be realized to distance $pR^{*} \approx R^{*}$ 
		(i.e. nearest-neighbor  hopping within the considered subnetwork). 
		So,  we take $p \equiv1 $ to maximize (\ref{phopRP}), 
		which   can be then expressed in the form 
\begin{equation}
P^{*}_{rp} \propto  \nu^{*}_{rp} e^{-2\alpha R^{*}}.
\label{hopRP}
\end{equation}
Taking into account that both parameters 
		$\nu^*_{rp}$ and $R^*$ 
		have finite and non-zero values 
		resulting from RP, and  
		considering that RP is not temperature dependent,
		it can be concluded that   
		electron hopping induced by RP is 
		temperature non-activated. 
%


The above  sketched  scenario reveals how RP 
		of metal ions of different valencies in the semiconductor lattice can
		{\em qualitatively} change nature of hopping process.
While the hop probability $P^{*}_{Mott}$
		given by proportionality  (\ref{Mott-vrhc}) for a system without RP
		is temperature-activated, 
		presence of RP causes that probability of hops
		$P^{*}_{rp}$ that dominate at lowest temperatures 
		is temperature non-activated. 
This unique property can be effectively used to explain metallic-like conductivity
 			of SmB$_6$ at lowest temperatures.  

Several experimental studies reveal that
		SmB$_6$ can be adequately treated as a system 
		with  IB in the forbidden gap, 
		and with the Fermi energy lying in the IB.
Because of nature of VFs, valence fluctuations between Sm$^{2+}$ and Sm$^{3+}$ 
	   will effect
		energies of localized impurity states similarly as RP in the model above,
		i.e.,
the VFs will cause time-dependent changes (fluctuations)
		of energy levels of impurity states.		
(See Fig.~\ref{fig2} and consider Sm$^{2+}/$Sm$^{3+}$ ions instead of $Me^{2+}/Me^{3+}$ ones.) 
So that, impurity states in SmB$_6$ are expected to be characterized 
   by energy fluctuating in intervals of non-zero width
		instead of energy levels
		with  energy constant in time.   
According to the above proposed scenario,
		and assuming that Fermi energy 
		lies in the impurity band of SmB$_6$, then
		 there will exist 
		impurity subnetwork consisting of impurity centers satisfying condition (\ref{Ef-band}).
Hence, mutual position of energies corresponding states in comparison with $E_F$
		is dynamically changing (fluctuating),
		while characteristic rate of these changes is closely related to a characteristic 
		rate VFs. 
This implicates favorable conditions
		for hops to empty states of less energy at high hopping rates, which can, in principle, 
		approach rates of VFs. 
So, based on (\ref{hopRP}), 
		which we consider as a relevant 
		case to discuss bulk properties of  SmB$_6$ [$R^{*} \ge R >>a$(SmB$_6$)], we can write
		probability of hops induced by VFs
		in the considered subnetwork, $P_{vf}^{*}$, as
\begin{equation}
P_{vf}^{*} \propto \nu_{vf}^{*} e^{-2\alpha R^{*}}.
	\label{hopVF}
 \end{equation}
Here $\nu_{vf}^{*}$ is defined analogously as $\nu_{rp}^{*}$, accepting the fact
		that VFs, in principle,  represent a special case of RP. 
Because both parameters $\nu_{vf}^{*}$ and $R^*$ 
		are governed by {\em dynamical} processes having origin in VFs (e.g. charge or spin fluctuations), 
		temperature dependence 
		of the probability of VFs-induced hops can be introduced 
		only via effect of temperature on these processes. 
However, if we take into account generally accepted fact that ground state of SmB$_6$
		is valence fluctuating state  \cite{Riseborough00,Wachter93,Antonov2002}, 
		then we have to conclude that parameters describing 
		dynamics of VFs (e.g. charge fluctuation rate, $\tau_{cf}^{-1}$) 
		do not converge to zero for $T \to 0$.  
>From this reason, $\nu_{vf}^{*}$ (which 
		is affected by {\em all} processes governing fluctuations of impurity energy levels),  
		can not be described by a temperature-activated law, 
		because in such case $\tau_{cf}^{-1}$ would have 
		to go to zero for $T \to 0$,
thus the ground state could not be a valence fluctuating state. 
Hence, the observed temperature non-activated conduction of SmB$_6$ at lowest temperatures
			can be reasonably explained by a temperature non-activated nature of dynamics of VFs, 
			which induces temperature non-activated 	
			hopping with probability described
		 by (\ref{hopVF}). 
>From this point of view,
		observation of almost constant resistivity in SmB$_6$
		at lowest temperatures indicates that dynamics of VFs in SmB$_6$
		is almost temperature independent for $T \to 0$. 
It can be also said that temperature dependent changes of dynamics of VFs,
		which consequently influence parameter  $\nu_{vf}^{*}$, should be directly
		reflected in temperature behavior of electric conduction,
		as follows from (\ref{hopVF}).
Thus, the proposed scenario not only gives possible explanation for 
		metallic-{\em like} resistivity behavior of SmB$_6$, 
		but also  gives new physical meaning and interpretation 
		to electrical conductivity data of SmB$_6$ and other 
		valence fluctuating semiconductors at temperatures approaching absolute zero temperature.

Here should be noted that because of the fact that electrical transport is realized via 			
		hopping process, corresponding electrical conductivity can be less 
		than the minimum metallic conductivity $\sigma_{min}$
		defined by Mott-Ioffe-Regel criterion. 
So, the proposed scenario naturally explains why 
		residual conductivity of SmB$_6$ 
		can be less than $\sigma_{min}$ 
		without need to consider some combination of 
		bulk semiconducting properties (obeying the criterion)
		and unusual metallic surface state (e.g. topologically protected).
Moreover, another surprising consequence of the model proposed here is that it 
		infers existence of enhanced surface conductivity 
		that can be even several orders of magnitude greater than the bulk one,
		what is in fact, qualitatively the same situation as expected
		for a Kondo topological insulator state in SmB$_6$.

Relatively higher surface conductivity of SmB$_6$ 
		can be according to the proposed model
		explained as follows.
Direct consequence of (\ref{hopVF}) is  
		that hopping probability 
	 exponentially decreases with increase of hop distance 	$R^{*}$.
Typical hop distance in bulk, $R^{*}_{bulk}$, 
		can differ from that of the surface, $R^{*}_{surf}$,
		if bulk and surface concentration of lattice imperfections 
		playing role of hopping centers are different.     
It is well known that low temperature properties of SmB$_6$ 
		change significantly from  sample to sample,
		indicating important role played by uncontrolled 
		defects, i.e. impurities 
		(typically $10^{-5}$ to $10^{-4}$ in atomic concentrations), 
		small departures from stoichiometry, and structural point 
		defects \cite{Nickerson1971,Morillo1983}.
Experimental results indicate that 
		a typical distance between lattice imperfections
		in high-quality SmB$_6$ 
		can be several orders of magnitude greater than the lattice parameter, 
		i.e. $R^{*}_{bulk} >> a($SmB$_6)$. 
However, situation has to be completely different  on the surface,
		because of much greater concentrations of lattice imperfections.  
The reason is not only ``more damaged'' surface layer, 
	 but also the fact that SmB$_6$ surface is ``covered'' by unterminated
		B- and Sm- bonds, which might play role of hopping centers. 
Considering $R^{*}_{surf} < R^{*}_{bulk}$, or even  $R^{*}_{surf} << R^{*}_{bulk}$, 
and taking into account proportionality 
		(\ref{hopVF}) it can be reasonably expected that hop probability in the surface layer 
		can reach values in orders of magnitude greater than those in the bulk.
Moreover, for completeness of the discussion it should be  mentioned 
		that because of relatively less values of $R^{*}_{surf}$ 
		and not excluded existence of localized surface states with a 
		localization length, $\alpha^{-1}_{surf}$, greater than one in bulk,
		it may arise situation that parameter $p=3 / 2 \alpha_{surf} R^{*}_{surf}$ 
		will satisfy the condition $p \ge 1$.  
In such hypothetical case	the value $3 / 2 \alpha_{surf} R^{*}_{surf}$
		 will maximize (\ref{phopRP}),
		 so that VFs-induced hop probability on the surface
		 will be not expressed by (\ref{hopVF}), but will be proportional to
\begin{equation}
		P^{*}_{surf} \propto  \nu^{*}_{vf}/(\alpha_{surf} R^{*}_{surf})^3,
\label{hopSVF}
\end{equation}
what however, indicates a different electrical transport regime.		 
Nevertheless, no matter whether hopping probability in the surface
		layer of SmB$_6$ is described by (\ref{hopVF}) or (\ref{hopSVF})
		(or even by combination of both of them because of presence of several types of localized
		surface states) it can be qualitatively concluded that hop probability
		in the surface layer is relatively greater than in bulk and it is temperature non-activated,
		i.e. metallic-like. 
Thus, the proposed scenario is qualitatively consistent with
		experimental studies of surface transport in SmB$_6$ \cite{Kim2012,Wolgast2013,Kim2013}
		which have demonstrated metallic surface and insulating bulk separation, 
		as well as studies of SmB$_6$ thin-films, which show
		systematic decrease of the resistivity and of the $\rho(4.2~K)/\rho(300~K)$ 
		ratio with decreasing thin-film thickness \cite{Batko1990}.   
It should be also emphasized that high surface conductivity of SmB$_6$ can be, in principle, 
		a result of combination of several types of conduction mechanisms in surface layer \cite{Chen2013},
    and here proposed VFs-induced hopping mechanism can be only one of them.

Similar scenario of temperature non-activated hopping type conduction 
		as we propose for SmB$_6$ 
		is expected to be present also in other semiconducting valence-fluctuating systems with IB(s)
		and Fermi energy lying in the IB.
For instance, qualitatively similar resistivity behavior to one of  SmB$_6$ 
		was observed also in FeSi \cite{Hunt94, Paschen97, Degiorgi95} 
		and YbB$_{12}$ \cite{Wachter93,Batkova2006}.
In fact, resistivity/conductivity saturation 
		at lowest temperatures should be a native property (a ``fingerprint") 
		of 		many ``real" VF semiconductors, 
		because impurity states (lattice imperfections)
		are as a rule introduced during the preparation process
		of these materials.

In summary, we have provided arguments that presence of valence fluctuations
		in semiconducting systems having localized in-gap states (impurity band) 
		introduced via lattice imperfections
		causes corresponding fluctuation of the energy of these states. 
Consequently, if the Fermi energy lies in the impurity band, 
		energies of in-gap states laying in a certain vicinity of the Fermi energy 
		fluctuate above and below Fermi energy,
	what creates favorable conditions for hops to empty states of less energy,
	i.e. without need of activation energy.
This give rise to temperature non-activated hopping process
		with the hop probability governed by dynamics of VFs 
		and concentration of hopping centers.
The proposed mechanism is in excellent qualitative agreement with many experimental observations
		indicating close relationship between concentration of lattice imperfections
		and residual metallic-like conductivity of SmB$_6$ at lowest temperatures,
		thus brings possible explanation of 
		unusual metallic like transport in  SmB$_6$ (and related Kondo insulators).
Moreover, because of the fact that (significantly) increased concentration of lattice
		imperfections that may play role of hopping centers is expected in the surface layer, 
		it predicts strong enhancement of surface conductivity in SmB$_6$.
In this sense the proposed scenario, which utilizes the well established concept 
		of  hopping conduction in solids,  
		enables to explain bulk and surface metallic-like properties
		of SmB$_6$ within common mechanism for bulk and surface conduction, 
		i.e. without need to consider ''unusual'' metallic surface conduction, 
		like existence of topologically protected surface. 
Nevertheless, question about quantitative conductivity estimation in SmB$_6$
		(both bulk and surface) due to VFs-induced 
		hopping process and a possible interplay of distinct types of surface conduction in SmB$_6$ 
		remains open and requires further studies.  

This work was supported by 
the Slovak Scientific Agency VEGA (Grant No. VEGA~2-0184-13).


\begin{thebibliography}{99}

\bibitem{Riseborough00}
P.~S. Riseborough, 
 \textit{Advances in Physics} \textbf{49,} 257 (2000).

\bibitem{Wachter93}
P. Wachter, in \emph{Intermediate Valence and Heavy Fermions},
  (Handbook on the Physics and Chemistry of Rare Earths,
  vol.19, North Holland,  1993).

\bibitem{Antonov2002}
V. N. Antonov, B. N. Harmon, and A. N. Yaresko, 
Phys. Rev. B  \textbf{66,} 165209 (2002).
    
\bibitem{Travaglini84}
T. Travaglini and P. Wachter, 
Phys. Rev. B  \textbf{29,} 893 (1984).

\bibitem{Namba1993}
T. Namba, H. Ohta, M. Motokawa, S. Kimura, S. Kunii, and T. Kasuya,
Physica B \textbf{186-188} 440 (1993).

\bibitem{Ohta1991}
H. Ohta, R. Tanaka, M. Motokawa, S. Kunii, and T. Kasuya,
J. Phys. Soc. Jpn. \textbf{50} 1361 (1991).

\bibitem{Alekseev1993}
P. A. Alekseev, V. N. Lazukov, R. Osborn, B. D. Rainford, I. P. Sadikov,
E. S. Konovalova, Yu. B. Paderno,
Europhys. Lett. \textbf{23,} 347 (1993). 

\bibitem{Frankowski1982}
L. Frankowski and P. Wachter,
Solid State Commun. \textbf{41,} 577 (1982).


\bibitem{Batkova2008}
M. Batkova, I. Batko, E. S. Konovalova, N. Shitsevalova, Y. Paderno,
Acta Phys. Pol. A \textbf{113} 255 (2008).

\bibitem{Batkova2006}
M. Batkova, I. Batko, E. S. Konovalova, N. Shitsevalova, Y. Paderno,
Physica B \textbf{378-380} 618 (2006).

\bibitem{Nickerson1971}
J.~C. Nickerson, R.~M. White, K.~N. Lee, R. Bachmann, T.~H. Geballe, and G.~W. Hull Jr., 
Phys. Rev. B  \textbf{3,} 2030 (1971). 

\bibitem{Allen79}
J.~W. Allen, B. Batlogg, and P. Wachter, 
Phys. Rev. B  \textbf{20,} 4807 (1979).

\bibitem{Batko93}
I. Bat$\!$'ko, 
P. Farka\v {s}ovsk\'y, K.  Flachbart, E.~S.  Konovalova,  and  Yu.~B. Paderno, 
Solid State Commun. \textbf{88,} 405 (1993).

\bibitem{Cooley95}
J.~C. Cooley, M.~C.  Aronson,  Z. Fisk,  and  P.~C. Canfield,   
Phys. Rev. Lett. \textbf{74,} 1629 (1995).

\bibitem{Flachbart01pb}
K. Flachbart,  S. Gab\'ani,  E. Konovalova,  Y. Paderno,  and  V.  Pavl\'ik, 
Physica B \textbf{293,} 417 (2001).

\bibitem{Gabani01}
S. Gab\'ani, 
K. Flachbart, E.  Konovalova,  M. Orend\'a\v {c}, Y.  Paderno,  V. Pavl\'ik,
 and  J.  \v {S}ebek, 
Solid State Commun. \textbf{117,} 641 (2001).
 
\bibitem{Gabani03}
S. Gab\'ani, E. Bauer, S. Berger, K. Flachbart, Y. Paderno,  
C. Paul, V. Pavl\'ik,  and  N.  Shitsevalova,  
Phys. Rev. B  \textbf{67,} 172406 (2003).


\bibitem{Tarascon1980}
J.~M. Tarascon, Y. Isikawa, B. Chevaliere, J. Etorneau, P. Hagenmuller, and M. Kasaya,
J. de Phys. \textbf{41,} 1141 (1980).

\bibitem{Cohen1970}
R. L. Cohen, M. Eibschutz, K. W. West, and E. Buller,
J. Appl. Phys. \textbf{41,} 898 (1970).

\bibitem{Pena1981}
O. Pe\~na, D. E. MacLaughlin, M. Lysak, and Z. Fisk,
J. Appl. Phys. \textbf{52,} 2152 (1981).

\bibitem{Zirngiebl86}
E. Zirngiebl, S.  Blumenroder, R.  Mock,   and  G.  G\"untherodt, 
JMMM \textbf{54-57,} 359 (1986).

\bibitem{Mock86}
R. Mock, E. Zirngiebl, B. Hillebrands, G. G\"untherodt,  and   F. Holtzberg,  
Phys. Rev. Lett.  \textbf{57,} 1040 (1986).

\bibitem{Fisk1996}
Z. Fisk, J. L. Sarrao, S. L. Cooper, P. Nyhus, G. S. Boehinger, A. Passner, 
and P. C. Canfield, 
Physica B \textbf{223-224} 409 (1996).

\bibitem{Caldwell07}
T. Caldwell, A.~P. Reyes, W.~G. Moulton, P.~H. Kuhns, M.~J.~R. Hoch,
P. Schlottmann, and Z. Fisk, 
Phys. Rev. B  \textbf{75,} 075106 (2007).

\bibitem{Miyazaki2012}
H. Miyazaki, T. Hajiri, T. Ito, S. Kunii, and S. Kimura, 
Phys. Rev. B \textbf{86,} 075105 (2012).


\bibitem{Edwards98}
P.~P. Edwards,  R.~L. Johnston, C.~N.~R. Rao, D.~P. Tunstall,  and  F. Hensel, 
Phil. Trans. R. Soc. Lond. A \textbf{356,} 5 (1998).

\bibitem{Ioffe60}
A.~F. Ioffe and  A.~R.  Regel,   
Prog. Semicond. \textbf{4,} 237 (1960).

\bibitem{MottDavis81}
N.~F. Mott  and E. Davis  in
\emph{Electronic Processes in Non-Crystalline Materials}
(Calendron Press, Oxford, 1971).

\bibitem{Dzero2010}
M. Dzero, M. Sun, K. Galitski, and  P. Coleman,   
Phys. Rev. Lett. \textbf{104,} 106408 (2010).

\bibitem{Dzero2012}
M. Dzero, M. Sun, P. Coleman , and  K. Galitski,   
Phys. Rev. B \textbf{85,} 045130 (2012).

\bibitem{Lu2013}
F. Lu, J.Z. Zhao, H. Weng, and  Xi. Dai,   
Phys. Rev. Lett. \textbf{110,} 096401 (2013).

\bibitem{Kim2012}
D. J. Kim, T. Grant, and  Z. Fisk,   
Phys. Rev. Lett. \textbf{109,} 096601 (2012).

\bibitem{Wolgast2013}
S. Wolgast, C. Kurdak, K. Sun, J. W. Allen, 
Dae-Jeong Kim,  and  Z. Fisk,   
Phys. Rev. B \textbf{88,} 180405(R) (2013).

\bibitem{Kim2013}
D. J. Kim, S. Thomas, T. Grant, J. Boltimer, Z. Fisk and JingXia,  
Sci. Rep. \textbf{3,} 3150 (2013).

\bibitem{Zhu2013}
Z.-H. Zhu,   
Phys. Rev. Lett. \textbf{111,} 216402 (2013).

\bibitem{Chen2013}
F. Chen, C. Shang, A. F. Wang, X. G. Luo, T. Wu, and X. H. Chen, arXiv:1309.2378v2.

\bibitem{Mott1968}
N.~F. Mott,
J. Non-Cryst. Solids, \textbf{1,} 1 (1968).

\bibitem{Shklovskij84}
B.~I. Shklovskii  and A.~L. Efros  in
  \emph{Electronic Properties of Doped Semiconductors},
  (Springer Series in Solid State Sciences, 1984).

\bibitem{Miller1960}
A. Miller, and E. Abrahams,
Phys. Rev. B  \textbf{4,} 2612 (1971).

\bibitem{Ambegaokar71}
V. Ambegaokar, B.~I. Halperin,  and  J.~S. Langer, 
Phys. Rev.   \textbf{120,} 745 (1960).


\bibitem{Morillo1983}
J. Morillo, C.H. de Novion, 
Solid State Commun. \textbf{48,} 315 (1983).

\bibitem{Pol-Sklov1991}
See A.R. Long in 
{\em Hopping Transport in Solids}, 
Modern Problems in Condensed Matter Sciences,
\textbf{28,} 207-231 (1991), and references therein.

\bibitem{Hunt94}
M.~B. Hunt, M.~A. Chernikov, E. Felder, H.~R. Ott, Z. Fisk,  and  P. Canfield,  
Phys. Rev. B \textbf{50,} 14933 (1994).

\bibitem{Paschen97}
S. Paschen, E.  Felder, M.~A.  Chernikov, L. Degiorgi, H. Schwer, H.~R. Ott, 
D.~P. Young, J.~L. Sarrao,  and  Z. Fisk,   
Phys. Rev. B \textbf{56,} 12916 (1997).

\bibitem{Degiorgi95}
L. Degiorgi, M. Hunt, H.~R. Ott,  and  Z. Fisk, 
Physica B \textbf{206-207,} 810 (1995).

\bibitem{Batko1990}
I. Batko, K. Flachbart, J. Mi\v skuf, V. M. Filipov,  E. S. Konovalova, Ju. B. Paderno,
J. Less-Common Met. \textbf{158} L17 (1990).

\end{thebibliography}

\end{document}